\documentclass[onecolumn]{emulateapj}

\RequirePackage{xspace}

\def\ifundefined#1{\expandafter\ifx\csname#1\endcsname\relax}

\ifundefined{ensuremath}\def\ensuremath#1{\relax\ifmmode{#1}}
\else${#1}$\fi\else\relax\fi
\ifundefined{nuc}\def\nuc#1#2{\relax\ifmmode{}^{#1}{\protect\mathrm{#2}}
\else${}^{#1}$#2\fi}\else\relax\fi

\shorttitle{Thermonuclear Supernovae: Probing Magnetic Fields by Late-Time IR Line Profiles}

\shortauthors{R. Penney \& P. Hoeflich}

\begin{document}

\title{Thermonuclear Supernovae: Probing Magnetic Fields by Positrons and Late-Time IR Line Profiles}

\author{R. Penney\altaffilmark{1,2}, P. Hoeflich\altaffilmark{1}}
\altaffiltext{1}{Florida State University, Department of Physics, Tallahassee, FL 32305, USA, phoeflich77@gmail.com}
\altaffiltext{2}{Clemson University, Tallahassee, FL 32305, USA, rpenney@g.clemson.edu}

\begin{abstract}

 We show the importance of $\gamma $ and positron transport for the formation of late-time spectra
in Type Ia Supernovae (SNe~Ia). 
 The goal is to study the imprint of magnetic fields ($B$) on late-time IR 
line profiles, particularly the [FeII] feature at  $1.644 \mu m$ which becomes  prominent two to three months after the explosion.
 As a benchmark, we use the explosion of a Chandrasekhar mass ($M_{Ch}$) White Dwarf (WD) and, specifically, 
 a delayed detonation model which  
can reproduce the light curves and spectra  for a Branch normal SNe~Ia. We assume  WDs 
with initial magnetic surface fields between 1 and $10^9 G$. We discuss 
large-scale dipole and small-scale magnetic fields.
 We show that positron transport effects must be taken into account for the interpretation of emission features
starting at about 1 to 2 years after maximum light, depending on the size of ${B}$.
 The [FeII] line profile and its evolution with time 
can be understood in terms of the overall energy input by radioactive decay and the transition
from a $\gamma$-ray to a positron dominated regime. We find that the [Fe II] line at $1.644 \mu m$ can be used
to analyze the overall chemical and density structure of the exploding WD up to day 200 without considering $B$.
At later times, positron transport and magnetic field effects become important.
 After about day 300, the line profile allows one to probe the size of the $B$ field.   
The profile becomes sensitive to the morphology of $B$ at about day 500. In the presence of a large-scale
dipole field, a broad line is produced in $M_{Ch}$ mass explosions which may appear flat-topped or 
rounded depending on the inclination at which the supernova is observed. Small or no directional dependence 
of the spectra is found for small-scale $B$.
 We note that narrow line profiles require central $^{56}Ni$ as shown in our previous 
studies. Persistent broad-line, flat-topped profiles require high density burning which is the signature of a WD close to $M_{Ch}$.   
Good time coverage is required to separate the effects of optical depth, the size 
 and morphology of $B$, and the aspect angle of the observer.
 The spectra require a resolution of about 500 km/sec and a signal to noise ratio of about
$20 \% $.  Two other strong  NIR spectral features  at about 1.5 and 1.8 $\mu m$ are used to demonstrate  
the importance of line blending which may invalidate a kinematic interpretation of emission lines.
 Flat-topped line profiles between 300 and 400 days have been observed and reported in literature.  
They lend support for $M_{Ch}$ mass explosions in at least some cases, and require magnetic fields
equal to or in excess of $10^{6} G$.
{ We briefly discuss the effects of the size and morphology of B on light curves, and limitations.
 We argue that line profiles are a more direct measurement of $B$ than LCs because they measure both the distribution of $^{56}Ni$
and the redistribution of the energy input by positrons rather than the total energy input.}
 Finally, we discuss possible mechanisms for the formation of high B-fields, and the limitations of our analysis.

\end{abstract}


\section{Introduction}\label{Introduction}
 Type Ia supernovae (SNe Ia) are invaluable fro probing the large-scale structure of the Universe and the origin of the elements. They
also provide important laboratories for the  study of the physics of flames, instabilities, radiation transport, non-equilibrium 
systems, and nuclear and high energy physics. 
 The consensus picture is that SNe~Ia result from a degenerate C/O white dwarf undergoing a 
thermonuclear runaway \citep{hf60}, and that these  in turn originate from  close binary stellar systems.
{\sl Potential progenitor systems} may either consist of two WDs, a so called double degenerate system (DD), 
and/or a single WD and a main sequence, Helium or Red Giant star, called a single degenerate system (SD).
Candidate progenitor systems have been observed for both the SD and DD cases \citep{greiner91,vdh92,rap94,kah97,
pilar04,gonz09,Kerzendorf09,Schaefer2012,Edwards2012}. For recent overviews, see proceedings of the IAU 281 Symposium edited by \citet{DiStephano13}. 

  Within this general picture for progenitors, two classes of {\sl explosion scenarios} have been discussed, which are distinguished by the triggering  mechanism of the thermonuclear explosion.
The first is the dynamical merging  of two C/O white dwarfs in a binary system.
 In this scenario, the thermonuclear explosion is triggered by the heat  of the merging process.
However, it is unclear whether the dynamical merging process leads to a SN~Ia, an ``Accretion Induced Collapse'',
or a WD with high a magnetic field. Another  problem is that there
seem to be too few potential progenitor systems \citep{webbink84,iben,benz90,rasio94,hk96,segretain97,yoon2007,WMC09,WCMH09,pakmor10,
loren09,isern11}. 

The second scenario involves the explosion of a C/O WD with a mass close to the
Chandrasekhar limit ($M_{Ch}$). The explosion in this scenario is triggered by compressional heating
near the WD center. Because the compressional heat released increases rapidly as the star nears $M_{Ch}$, the exploding stars 
are in a very narrow mass range \citep{hk96}.
  The donor star may be either a red giant or main sequence star of less than
7-8 solar masses or a helium star in an SD-system \citep{WI73}, or the accreted 
material may originate from a tidally disrupted WD in a DD-system \citep{WI73,Piersanti2004}.

 A comprehensive discussion of the progentors and explosion mechanisms is beyond the scope of this paper. For reviews and different views, 
 see \citep{branch93,pakmor10,ckjt13,gm13,h13}.
  Theoretical work and observational constraints from the spectra and light curves
favor the single degenerate scenario for the majority of cases,  with some contribution from the double degenerate scenario
\citep{hk96,sainom98,ww86_mergers,moch_livio90,sainom85,quimby06,shen_mergers11,h13}.                                             
  Within $M_{Ch}$ mass explosions, delayed-detonation models \citep{khokhlov91,woosley94,yamaoka92,gamezo03,gamezo04,p11}. Models
possessing a transition from a deflagration to a detonation front (DDT), have been found to reproduce the optical and infrared light curves 
and spectra of individual ``typical'' SNe~Ia reasonably well, including the time evolution 
\citep{h95,hk96,fisher98,nugent97,wheeler98,lentz01,marion09,maund10a,sim13,dessart14} and statistical properties, such as the small spread in the brightness-decline relation 
\citep{HKWPSH96,nughydro97,HGFS99by02,maeda03,kasen09,baron12}. The overall spherical structure in SN~Ia remnants and, in particular,
 the layered chemical structure observed in S-Andromeda are clear evidence of a detonation phase \citep{fesen07}.
 Though small, deviations from sphericity are to be expected due to rotation of the WD, off-center DDT and turbulence in the deflagration phase. These have been
shown to be present in  studies of spectropolarization, IR line profiles and supernovae remnants 
\citep{howell01,h04,motohara06,hoefl06b,gerardy07,fesen07,maund10a,fesen07,maeda11}. Recently, correlations between line profiles and light curves \citep{maeda11}
and line profiles and polarization \citep{maund10a} promise further progress towards a more complete picture of SNe~Ia.

 We regard four properties of SNe~Ia as the primary evidence in favor of
$M_{Ch}$ explosions:  A tight relation of the absolute brightness with the rate of the decline in brightness;
low continuum polarization \citep{wang97,howell01,patat12}; evidence for nuclear burning at densities $> 10^9 g/cm^3$
from late-time IR spectra \citep{h04,motohara06,maeda11} and narrow $Ni$ lines at late times as observed in SN 2003hv \citep{gerardy07}. 
In  SN 2003hv, late-time MIR spectra at about 130 days have been  obtained with the Spitzer Space Telescope. They show flat-top profiles consistent with the NIR profile and, thus, support the kinematic interpretation. 
Whereas the forbidden Co lines are broadly consistent with the expected distribution of $^{56}Ni$, 
late time $Ni$ lines are found to be narrow. Since $^{56}Ni $ has a decay time of about 6.3 days,
after several months only stable $Ni$ isotopes are left, and they exist only in the central region 
\citep{gerardy07,stritzinger14}.
Thus, at least some supernovae originate from $M_{Ch}$ mass explosion. 

 Despite the success of $M_{Ch}$ mass explosions,
whether they originate from SD or DD systems, there are serious problems within this picture
related to the mixing by Rayleigh-Taylor instabilities (R-T) inherent in
current  multi-dimensional simulations of deflagration fronts
\citep{khokhlov95,n95,livne99,rein99,gamezo03,plewa2007,roepke06}.
 During the deflagration phase, the unburned material is
heated by the diffusion of energy from the burned ashes to the unburned fuel. R-T instabilities
increase the surface area of the burning front and, thus, control the
rate of burning while mixing the burning products of different layers.
 Although a  layered chemical structure is partially restored during
 the detonation phase \citep{gamezo04,roepke12}, all current 3D models for the explosions predict R-T mixing
of iron-group plumes into the outer layers and mixing of the inner layers, both of which are at odds with observations.
 
 To understand the inconsistencies, it is necessary to both probe the distribution of radioactive material in SNe~Ia  and to look for
new physical effects not considered in current 3D simulations for deflagration fronts. Late-time line profiles of
[Fe II] and [Co III] probe
the distribution \citep{h04,motohara06,maeda11} and provided evidence for high magnetic fields in excess of 
3000 $G$ \citep{h04,sadler12}, a piece of physics which may influence the runaway and burning instabilities (see below). Moreover,
 advances in observational techniques have allowed measurement of spectra at late times
and thus are providing an increasingly comprehensive set of SNe~Ia data \citep{Phillips12,stritzinger14}.

The distribution of elements and  the strength and structure of $B$ may be a  key to a better understanding of SNe~Ia. However, depending on the phase, 
a direct translation of the Doppler shifts of a spectral feature is limited by optical depth effects,   
line blending, and a mix of geometrical asymmetries in the distribution. Moreover, a quantification of the $B$ fields requires 
detailed gamma and  positron transport. Past studies of line profiles were based on the assumption of local energy deposition of positrons.

The goal of this work is to present a systematic study of the effect of $B$ fields on the NIR line features. We use a spherical
density and chemical distributions and base our study on a DDT model as reference.
 We employ detailed Monte Carlo schemes for the transport for photons, gamma-rays 
and positrons.
In section 2, we will present a new positron transport scheme, and briefly summarize the 
other tools employed for the simulations. 
In the following sections, the results are presented and discussed.
 We will present a systematic study of the evolution of NIR profiles of [Fe II] and
[Co III] in the 1.6 to 1.7 $\mu m$ region. We show the importance of time-sequences to untangle 
the various effects.

\section{Methods}\label{Methods}

  Our analysis of the line profiles is based on the delayed detonation model 5p0z00.25, a 'Branch-normal' Type Ia which is part of the series published by \citet{HGFS99by02}. As in our previous 
analysis of the NIR  [Fe II] line \citep{h04}, the 3D-transport is calculated 
using a Monte Carlo scheme implemented in HYDRA, which includes detailed $\gamma $-ray
transport. Forbidden lines are included from databases published by \citet{kurucz93,kurucz02,NS88a,NS88b,LJS97,bowersetal97}.
 The new addition here involves the energy deposition by positrons and the extension from $\gamma$ to X-rays. 
These new modules will be described in the following.

\subsection{Positron Transport Scheme}

Several previous investigators have studied the transport of energy by high energy photons through supernova envelopes 
\citep{as88,burrows90,chan91,1994Hoeflichgamma,Hoeflich:gamma2002}. 
 As part of his PhD thesis,  new bound-free and bound-bound transitions have been implemented which 
 allows us to calculate X-ray spectra \citep{penney11}. This transport module has been further extended
to incorporate positron transport in the presence of magnetic fields.

Several authors have addressed the transport of positrons \citep{brd79,bl87}.  Early studies by  \citet{colgate80} and \citet{lingenfelter93} used one-dimensional
approximations and assumed the magnetic field was either chaotically twisted at a scale small enough to trap the positrons in place or radially combed by the expansion, 
restricting the positrons to move radially. They calculated the energy deposition by positrons at late times and demonstrated  the extent to which positron escape would
lead to light-curves (LCs) declining faster than the rates of the radioactive decay driving them.   
  \citet{milne99} presented calculations using cross-sections and a Monte-Carlo technique more similar to ours, but continued to use 
either a  radial magnetic field or ones which are chaotic and sufficiently large to trap all positrons. As part of the thesis of \citet{penney11}, light curve results have been 
compared to those published \citet{milne01}. { As discussed below, we found a  generally good agreement,  but with a slightly reduced positron escape.
 In this paper, we will make use of line profiles to measure positron transport effects
and their dependency on $B$. Profiles can be expected to more sensitive and 'cleaner' than LCs because they measure directly the change in 
the redistribution function of the energy deposition from radioactive decay. In contrast, LCs rely on one quantity, the total energy deposition in the
envelope that powers the bolometric LC. The reconstruction of the bolometric LC depends in turn on the photon redistribution in frequency space and, after a few years,
other energy sources may become important such as interaction between the SN and the interstellar material (ISM).}

 \subsection{Positron Creation and Cross Sections}
 The primary source of positrons is the $\beta ^+$ channel of the $^{56}Co \rightarrow ^{56}Fe^* $ which accounts for about $18\%$ of all $^{56}Co$ decays.
About $1.4 MeV$ of the total excitation energy of $^{56}Fe^*$ is { available to be split between neutrinos and positrons }, with the resulting positron energy spectrum  \citep{nadyozhin} given by  
\begin{equation}\label{eq:trans_spectrum}
N(E)=Cp^2(E_o-E)^2(2\pi\eta(1-\exp(-2\pi\eta))^{-1})
\end{equation}  where $C$ is a scale factor and $\eta$ is the charge of the nucleus times $\hbar$ over the velocity of the electron.  The mean energy of the spectrum is .632 MeV.  

A small number of relatively high energy positrons are also contributed by pair-production 
from the high-energy gamma-ray lines of $^{56}Co$ decay.  As most $\gamma $s escape without 
interaction at late times, these contribute only a few percent to the total positron 
production by 300 days, but their production and transport is included.  The cross-section for this channel of  positron  production has been  discussed by \citet{amba88}. 

For the Monte Carlo transport, we take into account three dominant processes of interaction: scattering off atomic electrons as the positron moves through the media, the similar excitation of free electrons in a plasma, and direct annihilation with electrons.  

\begin{figure}
\begin{center}
\includegraphics[angle=270,width=0.68\textwidth]{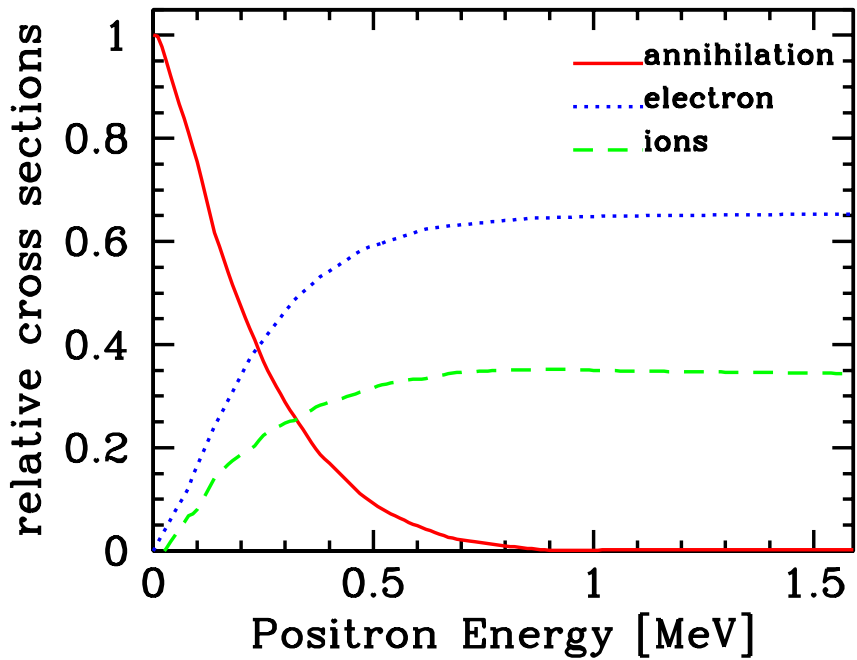}
\end{center}
 \caption{The strength of the electron, plasma, and annihilation cross sections
for a single ionized plasma. Annihilation dominates at low energies after the
positrons lose most of their energy by electron and plasma interaction.
 Electrons dominate plasma interactions for low ionization levels but ions will
dominate in highly ionized plasmas.}
\label{pcross}
\end{figure}  

The annihilation  cross-section is given by  eq. \ref{eq:electronsigma} \citep{astroformula}:
\begin{equation}\label{eq:electronsigma}
\sigma = \frac{\pi r_o}{\gamma+1}[\frac{\gamma^2+4\gamma+1}{\gamma^2-1}ln(\gamma+\sqrt{\gamma^2-1})-\frac{\gamma+3}{\sqrt{\gamma^2-1}}]
\end{equation} with $r_o=e^2/mc^2$ and $\gamma$ is the Lorentz factor. 
In undergoing this interaction, the positron is destroyed, its kinetic energy is deposited into the surrounding material and its rest-mass energy is released as gamma-rays.  

Following  \citet{lingenfelter93} and \citet{gouldsigma}, the energy loss by interaction with  charged particles is given by
\begin{equation}\label{eq:bloch}
\frac{dE}{dx} = \frac{-4\pi r_o^2m_ec^2q}{AM_n\beta^2}(qln(\frac{\sqrt{\gamma-1}\gamma \beta m_ec^2}{b_{max}})\Pi (\gamma)
\end{equation}
$\Pi(\gamma)$ is a relativistic correction given by 
\begin{equation}
\Pi (\gamma) = \frac{\beta^2}{12}[\frac{23}{2}+\frac{7}{\gamma+1}+\frac{5}{(\gamma+1)^2}+\frac{2}{(\gamma+1)^3}]
\end{equation}
where $q$ is the relative charge of the particles in the media. For atomic electron interactions, $q$ is 
the atomic number $Z$ for the atom. For plasma scattering, $q$ is the ionization fraction.
$b_{max}$ is the maximum impact parameter, equivalent to the maximum amount of energy the positron can loose in
 one interaction.  For atomic electron scattering, it is the the ionization potential.  
 Since the dependence on this parameter is weak, we simplify by using an average ionization potential for the medium. 
For the case of interactions with electrons in the plasma, the maximum impact parameter is $\hbar \omega$ 
because the impact with a free electron sets up a disturbance in the plasma which has an energy proportional to the frequency.
  Note that the ionization stage depends on time and requires ionization models which, in turn,
will depend on the energy deposition by positrons. 
In this study, we assume single ionization.  Other sources of continuous energy loss, those of Bremsstrahlung and synchrotron radiation, were
 considered but found to be minor contributions and, for computational efficiency,  have been omitted in this study.

The relative strengths of the three types of cross-sections are shown in Fig. \ref{pcross}.  
Annihilation is a very small part of the cross-section through-out most of the energy range of the 
positron spectrum.   As the positrons scatter and continuously loose energy, though, they will either fall 
into the range where direct annihilation  dominates and they be destroyed, or they will slow down until their energies  become similar to the binding energy of atomic electrons, form positronium, and 
be annihilated.  

Between 1 KeV and the binding energy of nearby electrons, the positron can undergo charge exchange, stripping an electron off nearby atoms to form positronium, which has a negligible lifetime before undergoing either 
$e^+e^- \rightarrow 2\gamma$  or $e^+e^- \rightarrow 3\gamma$ for para- and ortho-positronium, respectively. We assume that the para- and ortho-state of positronium are in statistical equilibrium  
 for the creation of $\gamma $ photons. The cross-section of this process for elements other then H is not well known, but is generally proportional to the Bohr-radius, and thus much larger then the direct annihilation cross-section.  

This uncertainty in the cross-section for annihilation at low energies will hardly influence the location of the energy deposition because the energy loss
is large (see  eq. \ref{eq:bloch}).
Even at day 1000 after the explosion, the range of a 1 KeV positron is less than 0.1 \% of the radius of the envelope.  
Thus, we assume that once a positron has fallen below 1 KeV, it deposits its 
kinetic energy locally and annihilates via positronium formation.  
 \citet{lingenfelter93} found that some
 small fraction of thermalized positrons ($ \approx .1\% $ in models similar to ours) will avoid annihilation , 
 surviving as ''slow'' positrons as the envelope density becomes low enough 
 for the lifetime before positronium formation to become very long.  However, as 
 these will have already deposited their kinetic energy in the ejecta, and their mass-energy would not 
 contribute whether they annihilate or not. At late times,  $\gamma $s from annihilation are almost 
 guaranteed to escape the envelope without interaction.

\subsection{Positron Transport:}
The transport is solved via a Monte Carlo method very similar to the photon transport \citep{1994Hoeflichgamma,Hoeflich:gamma2002}.
A background model is discretized and the radiation transport is performed in individual computational cells.
Upon interaction, the process is chosen randomly, weighted by the size of the individual cross-sections.

The motion of positrons depends on the presence, size and morphology of the magnetic field $B$ imposing 
a Lorentz force on the positron. The positron gyros around the magnetic field line. The scale imposed 
by $B$ is the Lamour radius
\begin{equation}\label{eq:Lamour}
r_L = \frac{p_{\perp}}{qB}     \mbox{     ~($p_{\perp}$ the momentum perpendicular to the field.)}
\end{equation}
 The significance of the gyro-motion depends on the size of $r_L$ relative to the scale of variations 
of the explosion model $r_M$, such as the density scale height or the scale of the $B$ field changes.

  For computational efficiency, we distinguish three different implementations: 1) $r_L >> r_M$, 2) $r_L << r_M$, 
and 3) the case in between.

 In case 3, the path of the positron must be explicitly integrated by breaking the path-length into segments 
sufficiently small so that the B-field changes little along individual segments. A positron propagates along a circular arc 
on that path (the radius determined by eq. \ref{eq:Lamour}) until  
it is absorbed, scatters, or escapes the envelope. 

Case 1 and 2 avoid the need to reconstruct the path for individual positrons and, thus, they allow for a higher computational
efficiency by a factor of more than $\approx 10$.

 For case 1, small magnetic fields, the path of the positron is nearly straight and we integrate along 
linear rays until the positron is absorbed, scattered or escapes. 

\begin{table}
\begin{center}
\caption{Maximum Lamor radius as a fraction of the radius of the supernova envelope at various times  and initial
 surface field strengths. Here, the radius is defined as the layer expanding with a velocity of 40,000 km/sec. Typically, turbulent velocities in the
progenitor and during the deflagration phase of burning are between $\approx 200$ to $\approx 1000  km/sec$ \citep{hs02,gamezo04}, or $5 \times 10^{-3}$ and $2.5 \times 10^{-2}$ in units of the fractions given.}
\begin{tabular}{|c|c|c|c|c|c|}
\hline field/time & 20d & 60d & 100d & 300d & 500 d\\ 
\hline $10^3 G$ & .38 & 1.15 & 1.9 & 5.7 & 9.6 \\ 
\hline $10^4$G & .03 & .11 & .19 & .57 & .96 \\ 
\hline $10^6$G & .003 & .001 & .001 & .005 & .009 \\ 
\hline $10^9$G & $3E{-6}$ & $1.1E{-6}$ & $1.9E{-6}$ & $5E{-6}$  & $9E{-6}$ \\ 
\hline 
\end{tabular}
\label{table1}
\end{center}
\end{table}

 For large magnetic fields, $r_L$ is small compared to the structure of the explosion model. The positron can be
assumed to follow the field line on a spiral trajectory
mean-free path. 
The general motion of a  positron is along the field line with its path length increased by a factor of $1/sin(\theta_p)$, where 
$\theta_p$ is the pitch angle the positrons momentum vector makes with the field line. Effects of gradients in $B$ are implemented
following \citet{chenplasma}.  We divide these into two forces, one parallel  to the field lines, and one perpendicular.
The former causes the ''magnetic mirror'' effect. Because the first adiabatic invariant is constant, a change in the B field strength along 
the direction of travel causes a change in the pitch angle.
 In the code, we average the derivative of the B field over the zone in the direction of the field lines, and use this to calculate an average gyro-radius 
for that zone. The result is that positron diffusion is discouraged in the direction of increasing magnetic field and encouraged in 
the opposite direction, as an increasing pitch angle will increase the pathlength. 
  The effect of the perpendicular component of the $B$ gradient causes the  center of the gyro-radius off the field line
due to gradients perpendicular to the $B$ field.    
 We add this velocity to the motion of the particle in each zone using the following equation
\begin{equation}\label{eq:driftperp}
v=\frac{1}{2}v_{\perp}r_L\frac{B \times \nabla{B}}{B^2}
\end{equation}
 
 For illustration, table \ref{table1} shows the evolution of $r_L$ for a typical SNIa model as a  
function of time. In practice, we use cases 3 or 1 if $r_L$ is smaller than the resolution of the
background model or larger than twice the model grid, respectively. Cases 1 and 3 have been tested
against the general case 2.
 With time and lower magnetic fields, the range of the positron transport increases. 
For $B$ fields larger than $10^{9}G$, positrons follow the field lines. For $B$ fields less than
$10^3 G$ and about one year, the bending of the path of the positrons remains small and 
they travel along rays. We note that turbulence will produce structures of about $1$ to $5  \% $ of
the envelope structure. Thus, the energy deposition by positrons requires full transport of positrons
for all $B$ fields less than $B \approx 10^6 G$ when the transport effects are important.

\section{The Reference Model}
\begin{figure}
$\begin{array}{cc}
\includegraphics[width=0.45\textwidth]{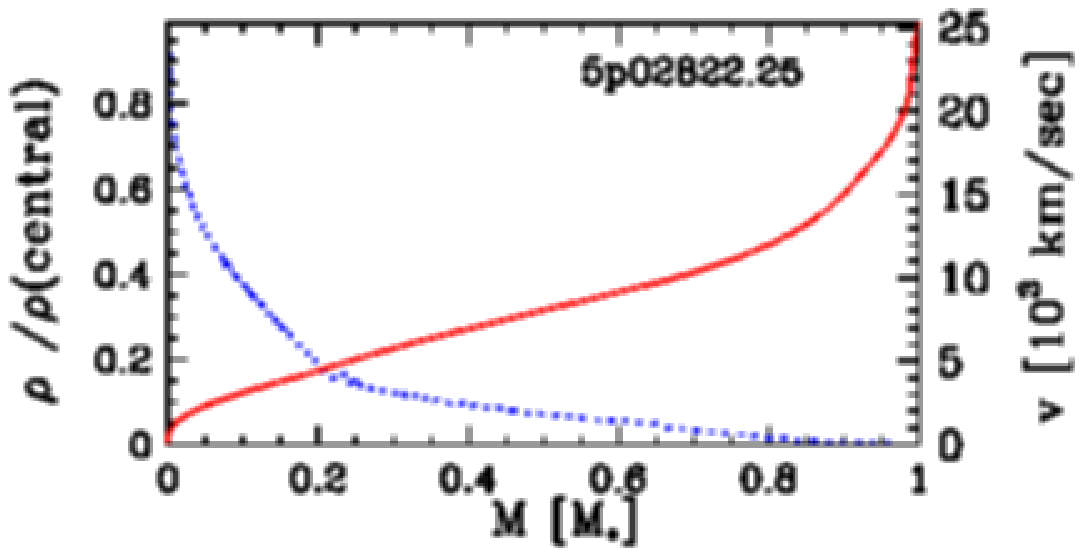}  &
\includegraphics[width=0.46\textwidth]{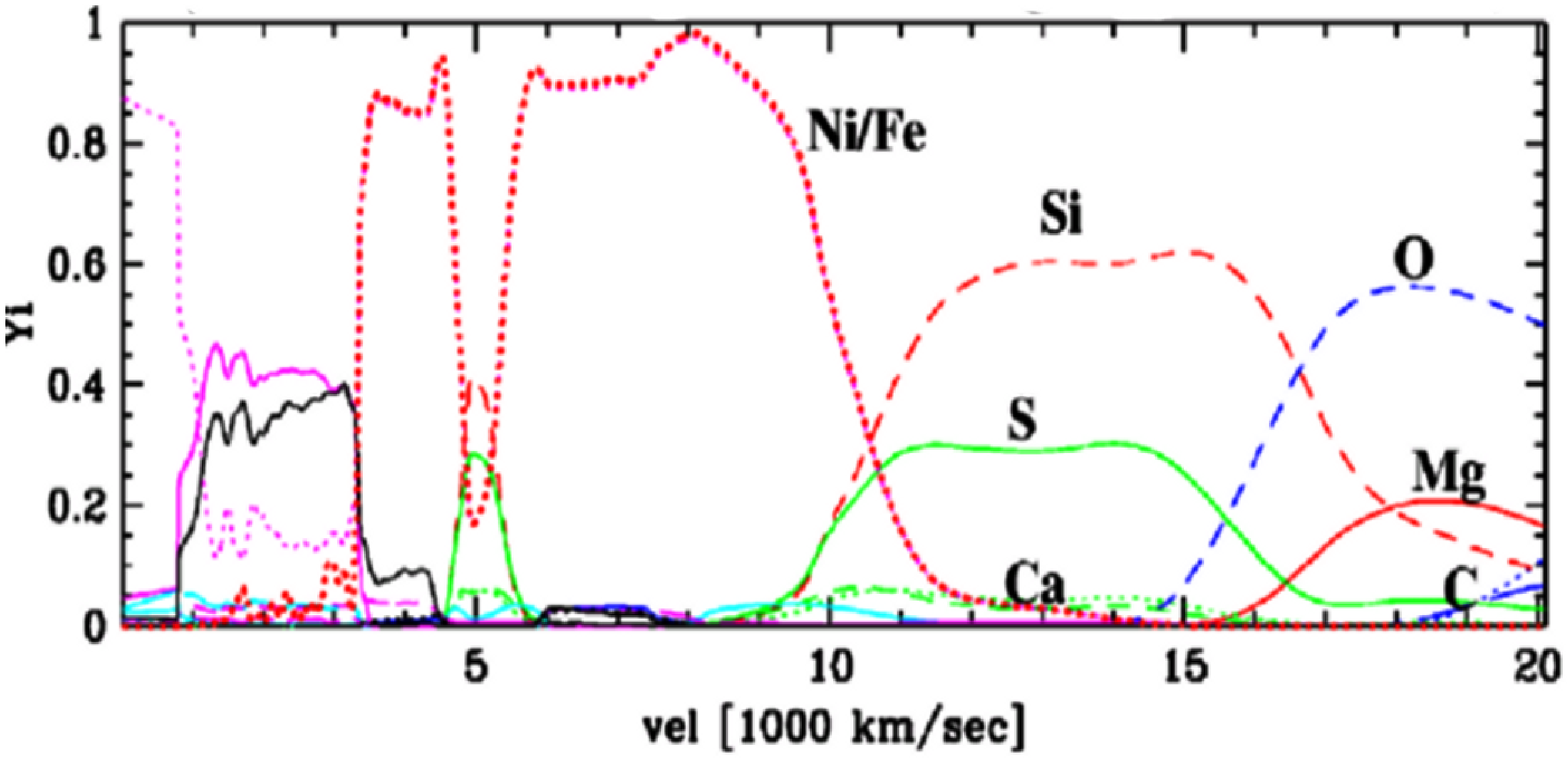}
\end{array}{c}$
\caption{{ Structure of a spherical delayed detonation model which can reproduce LCs and spectra of Branch-normal SNe~Ia \citep{HGFS99by02}.}
Density (blue, dotted) and velocity (red, solid) as a function of the mass are given on the left. Abundances of the most abundant stable isotopes as a function of the expansion velocity are given on the right. In the center,
the abundances correspond to  $^{54}Fe, ^{58}Ni$ and $Co$.
We note that current 3D calculations predict similar density structures but with strong chemical mixing.}
\label{chem} 
\end{figure}

 This study uses a spherical, delayed detonation
model as a reference model.  This model has been successful in reproducing
the optical and IR light curves and spectra of a Branch normal SNe~Ia, 5p0z22.25 \citep{HGFS99by02}.
 The C/O white dwarf originates from a progenitor with a   
main sequence mass of  $5M_\odot$ and solar metallicity.  
At the time of the explosion, the central density of
the WD was  $2 \times 10^9 g/cm^3$, and the transition
from deflagration to detonation was triggered at a 
density of $25 \times 10^6 g/cm^3$. 
 About 0.6 $M_\odot$ of $^{56}Ni$ is produced.
 The density, velocity and chemical  structure is given in Fig.\ref{chem}.
As typical for $M_{Ch}$ mass explosions,
the high density burning in the central region produced stable isotopes
of iron-group elements rather than radioactive $^{56}Ni$.
 Spherical models suppress mixing by Rayleigh-Taylor instabilities  and, as a consequence, 
we produce a central 'hole' in the $^{56}Ni $ distribution  of about $3000~km/sec$.
 The spherical explosion model and evolution has been calculated using 912 radial zones in 
the comoving frame. Positron transport is inherently 3-dimensional in presence of $B$ fields.
For the $\gamma $ and positron and NIR transport, the model has been remapped 
into the observers frame using an Euclidian grid with 330 radial zones with a resolution of about 50 km/sec up to the
Si-rich layers. For $\theta $ and $\phi$, we use 70 zones with equidistant spacing in  $sin~\theta $ and $\phi$.
 The energy deposition has been stored on a Cartesian grid of $501^3$
with a matching resolution of $100 km/sec$. The emitted photons have been detected in 21 counters each 
both in $\theta $ and $\phi$, and about 2500 counters in frequency/energy. The averaged positron spectrum is 
detected in 100 counters. A typical run uses a package of $1 - 5~10^{10}$ photons and positrons.  

To study the effects of $B$, we imprint various fields onto the model described above.
We assume that the initial field prior to the explosion of the WD can be described as a dipole with a surface averaged 
field of $B$.  For the rotation, we assume that the WD is a  rigid rotator with angular velocity $w$, and a constant magnetization $m$.
  Taking the field due to a thin  spherical shell and summing to the radius of the dwarf $R_o$ (taking R to be the radius of the White Dwarf at 90\% of the 
 Chandrasekhar Mass, 1600 km), the structure of the  field is given by
\begin{equation}\label{eq:magfield}
\vec{B_o}(r,\theta) = \omega m[(1-\frac{3r^2}{5R^2_o})\cos(\theta)\hat{r} - (1-\frac{6r^2}{5R^2_o}\sin(\theta))\hat{\theta}]
\end{equation}

$B$ field strengths are considered between 1 and $10^9G$. With the surface field as boundary 
condition, we use equation \ref{eq:magfield} to calculate  the magnetic field throughout the progenitor.  Note that the maximal range 
of the field strength is larger by about a factor of 4 close to the center of the initial WD.
 We can assume that the magnetic field is frozen into the matter during the explosion.  
 The final structure of the field is very similar because of the similarity of the 
initial structure and the homologous expansion of the WD.
After the initial acceleration phase, a few minutes into the expansion, the time evolution of the field at every point decreases as the
 square of the radius and thus, with time because of conservation of the magnetic flux. 
 During the homologous expansion field strength at $r$ is given by $B_o(\frac{r_o}{r})^2$.

\section{Results}\label{Results}

 In this section,  we will discuss the use of IR lines as tool for studying the underlying chemical structure
and the signature of the strength and structure of the magnetic fields at late times. { The use of LCs is also considered briefly, and compared to that of IR lines}.
 Both light curves and spectral line profiles allow the probing
of magnetic fields. The advantage of broad-band light curves are that they provide high photon statistics. This allows one to obtain accurate
LCs until very late-times and, in principle, for a large number of objects. However, late time LCs contain limited information 
because they mostly depend on one quantity only: the combined escape probability of $\gamma-$ rays and positrons. 
Spectra contain information about the escape probability and the time-dependent redistribution functions but limiting
photon statistics put high demands on the observations and limits their use to 'local' supernovae.
 Both the resulting LCs and spectra can be understood in terms of the energy deposition.

\subsection{Energy Deposition by X- and $\gamma$-Rays and Positrons}

 Radioactive decay of $^{56}Ni \rightarrow ^{56}Co \rightarrow ^{56}Fe $
is the dominant energy source which powers the light curves and spectra of SNe~Ia.
 The total energy input depends on the total mass of the initial $^{56}Ni$
and the redistribution by transport effects.
 $^{56}Ni$ decays by electron capture which leads to the emission of
high energy photons when the excited $^{56}Co$ transitions to its ground state.
 Besides decay via electron capture, $18 \% $ of all decays of $^{56}Co$
are $\beta^+$ and lead to the emission of a positron. 

\begin{figure}
\includegraphics[width=0.98\textwidth]{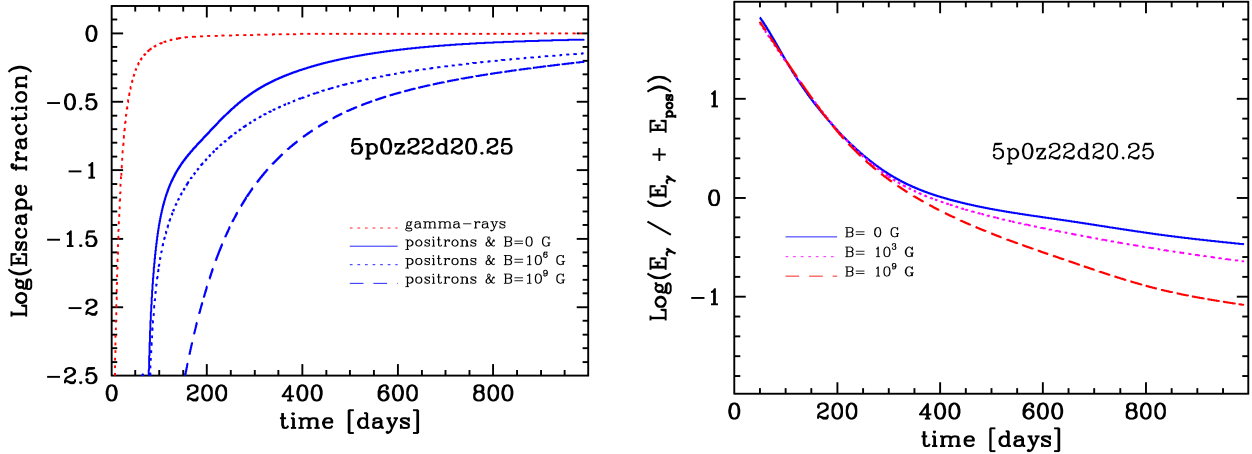} 
\caption{Escape probability for photons and positrons (left) 
and contribution to the total energy input in \% (right) as functions of time.
Magnetic fields influence positron transport only. Probing magnetic fields
requires late-time observations.
 }
\label{esca}
\end{figure}

\noindent
{\bf Escape Probabilities:} First, we want to discuss the role of the escape probabilities 
and their implications for LCs (Fig. \ref{esca}). 
The small cross sections for $\gamma-$ rays result in rapidly increasing escape fractions. 
 Already at maximum light, $\approx $ 20 days after the explosion, 15 \% of all $\gamma$s escape. 
 This fraction increases to about 85 and 98 \% by day 100 and 300, respectively. 
 In contrast, the large positron cross sections cause almost complete local annihilation of positrons 
up to about 200 days. The length of the positron path depends on the size of $B$, which 
causes a strong dependence of the positron escape fraction on $B$ at later times.
 However, as the kinetic energy of positrons only account for some 3\% of the total energy produced by $^{56}Co$ decay, 
 the total energy input is dominated by $\gamma $ rays until around day 150-200 (Fig. \ref{esca}). 
Thus studying the $B$ field via the influence of energy deposition on optical/IR wavelengths inherently requires late-time observations.

\begin{figure}
\begin{center}
$\begin{array}{c}
\includegraphics[width=0.78\textwidth]{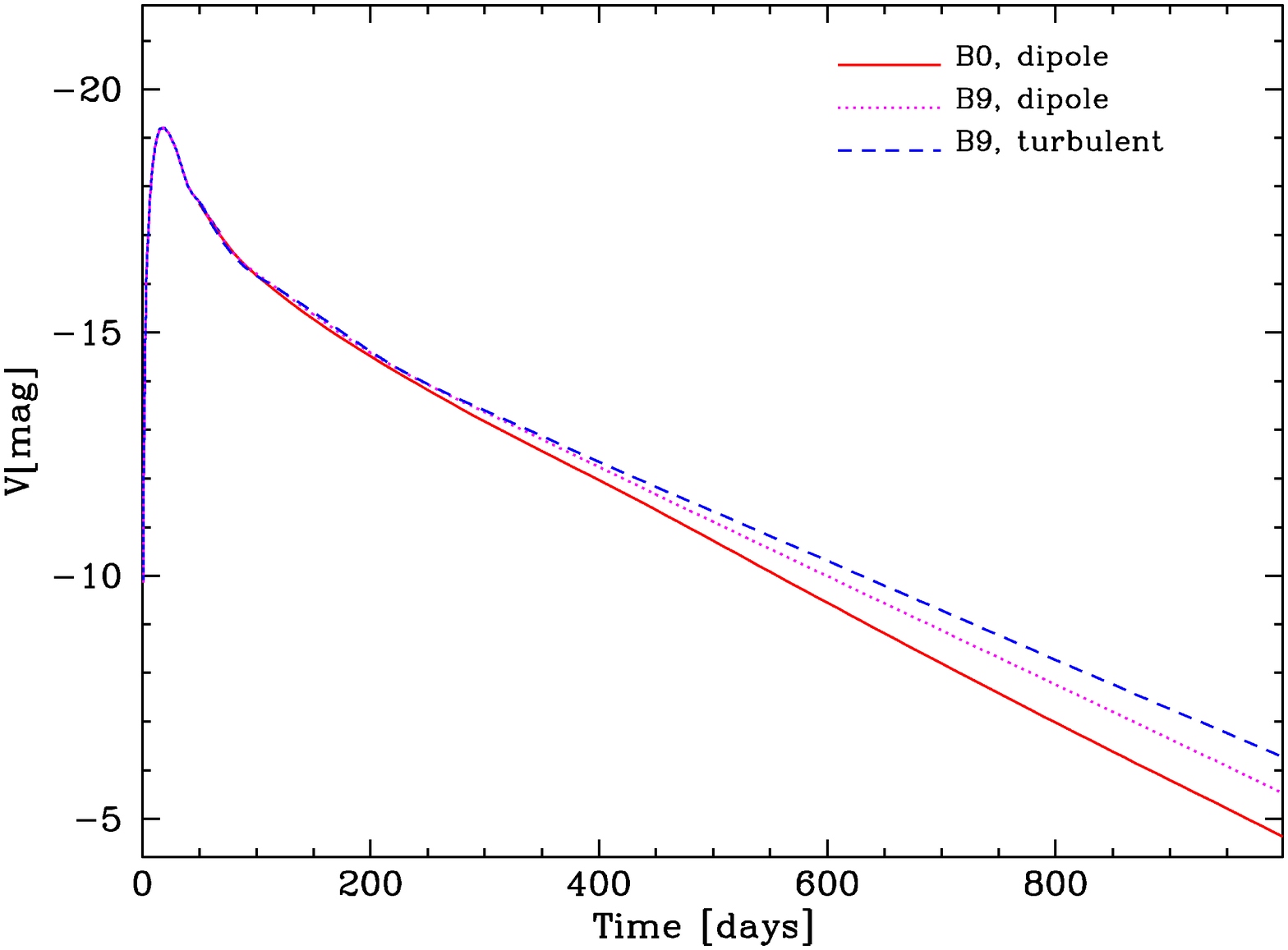} 
\end{array}$
\end{center}
\caption{Influence of the size and morphology of $B$ on the visual light curves for our reference model. V is given for  dipole 
and turbulent fields.
}
\label{lc}
\end{figure}

\noindent
{\bf Light Curves:}
{Previously, detailed simulations for positron transport effects on the bolometric light curves have been published by \citep{milne99} 
for a wide range of models based on the deflagration model W7 \citep{nomoto84}, the delayed-detonation models \citep{yamaoka92,hk96,hwt98}, helium detonations  
\citep{Ruiz93,hk96} and pulsating-delayed detonation and merger models \citep{hkw95,hk96}. We have compared the energy escape by positrons 
of our reference model and dipole fields with  model DD23C \citep{hwt98} and  radial magnetic fields  used in the sample of \citet{milne99}.
 We find that our our escape fractions agree with \citet{milne99} within $\approx 20 \% $. For high $B$ fields, our escape fractions 
are slightly smaller because we include non-radial fields whereas \citet{milne99} consider the radial component only.

 For a brief discussion, we use the V-Band (V) as a proxy for the bolometric LC \citep{arnett80}.
LCs are shown in Fig. \ref{lc} for low and high $B$ for dipole and turbulent fields.
The scale of the turbulent field has been taken to be one pressure scale height in the progenitor WD.
 The formation of early-time optical and IR light curves have been previously discussed \citep{hwt98,HGFS99by02,dessart14}.}

 Between 60 and 200 days, the decay of V is steeper than is expected from the decay of $^{56}Co$
because $\gamma $-ray photons dominate the energy input and they rapidly increase their escape probability, as shown in Fig. \ref{esca}.
 In fact, this discrepancy caused early predictions that SNe~Ia are not powered by $^{56}Ni$ \citep{Colgate62}.
Note that the decline rate increases at about 100 to 200 days with decreasing brightness because more centrally concentrated $^{56}Ni$
causes a delay in the $\gamma $ escape \citep{hmk93a,HGFS99by02}.

After 300+ days, almost all $\gamma$ rays escape and the LC is dominated by positrons. As the envelope is now optically thin
for optical photons the anisotropy effects are expected to be small.
 At $\approx 300 days$, V  closely mimics the decay curve of cobalt. { The change of slope after this point may be used to estimate
$B$ \citep{milne99}. At about 300 days, the influence of $B$ remains small $\approx 0.05 ^m$. After about 3 years, the influence of $B$ grows to  
 about one magnitude. Even for high $B$, the path of positrons follows the magnetic field line allowing some positron mobility and, then, positron escape.
 In contrast, for a turbulent field photons are trapped, which results in local energy deposition (see Fig. \ref{lc}) until the field weakens to the point that the Lamour radius is larger than the size of the turbulence.
 Despite the significance of both  the size and morphology of the $B$ field on late-time bolometric light curves, there are several limitations in using LCs to probe $B$.
 Monochromatic light curves are needed to reconstruct bolometric LC. However, V or UBVRI may not actually represent the bolometric light curves.
 It  has been suggested that emission will shift from the V-Band to the infrared as the timescales of atomic 
transitions in the optical range increase relative to those in the infrared due to falling matter densities 
(called the 'IR catastrophe'). In models, the 'IR catastrophe' may set in anywhere from 300 to 400 days depending on the details of the atomic
model and data, and the abundances \citep{ircatastrophe}. Moreover, recent observations of SNe~Ia identified mid-IR emission lines as 
major cooler starting  at about 3-4 month after the explosion \citep{gerardy07,Li14}. Finally, other energy sources may contribute to the energetics, such as interaction of the envelope with its surroundings,
which will deposit energy in the high velocity layers. Moreover, the diversity among SNe~Ia has been well established 
and variations in the $^{56}Ni$ distributions will change the bolometric light curves.
  These uncertainties and variables make it difficult to extract $B$ field strengths from just the one quantity provided by LCs.}

\begin{figure}
$\begin{array}{cc}
\includegraphics[angle=270,width=0.33\textwidth]{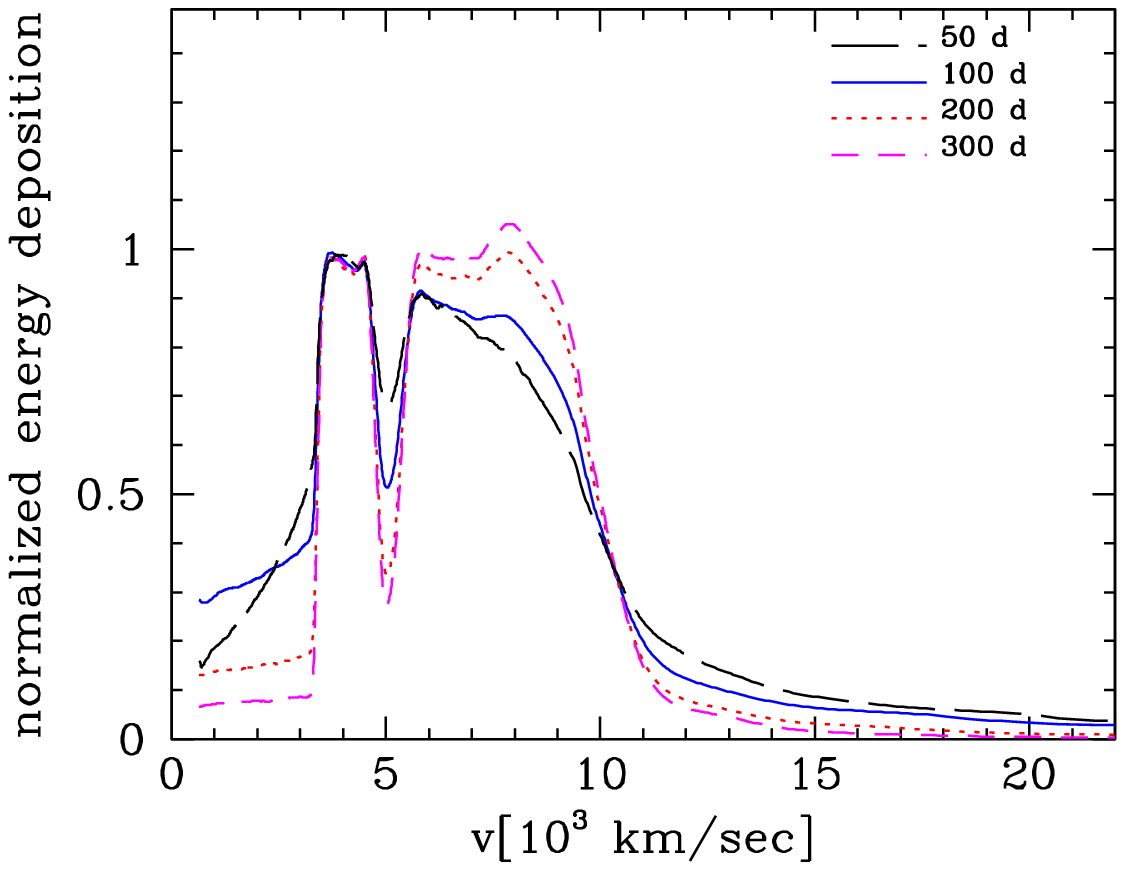} &
\includegraphics[angle=270,width=0.66\textwidth]{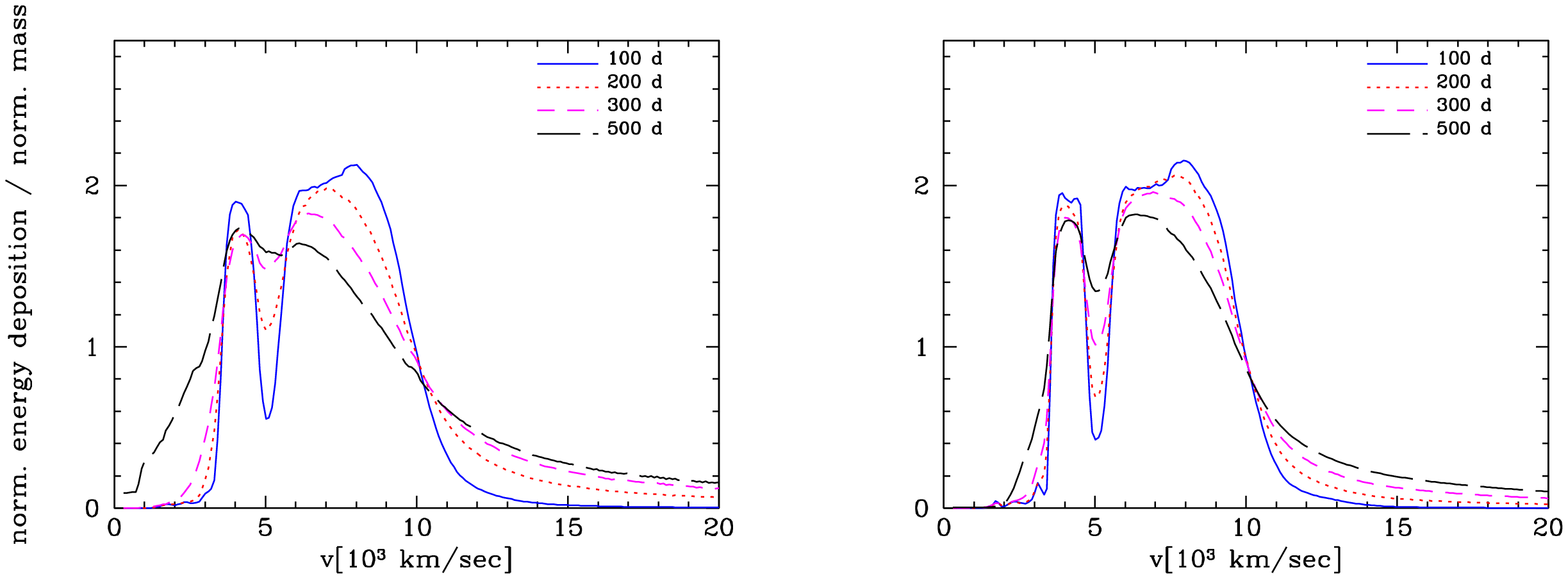}  
\end{array}$
\caption{Energy input $\dot{E}$ per {\bf gram and second} by both $\gamma $-rays and positrons assuming local trapping (left), 
and angle-averaged by positrons for initial magnetic fields of $B=1 $ (middle) and $10^9 G$ (left), respectively.
 We normalized the total  $\dot {E}$ to $ v \approx 4000 km/sec $  to compare the overall distributions and to
compensate for the rapid decrease of trapped energy by $\gamma $-rays (see Fig. \ref{esca}).
$\dot{E}$ of positrons has been normalized {\bf to the average over the entire envelope}. 
}
\label{epositrons}
\end{figure}

\noindent
{\bf The Energy Distribution:}
We want to shift our discussion to the distribution of the energy deposition and the resulting line profiles.
 The energy depositions by  $\gamma $-rays and positrons  and positrons only are given in Fig. \ref{epositrons}. 
Up to about  100 to 200 days, the contribution from positrons is locally deposited due to high cross sections. Non-locality is produced by $\gamma$-rays.
At early times, about a week after the explosion, the optical depth for $\gamma $-rays is still large and the energy input 
closely traces the distribution of the radioactive isotopes (Fig. \ref{chem}).
We can see the ``hole'' in the center occupied by stable, electron capture elements.  Due
 to geometrical dilution, the optical depth for $\gamma $
photons decreases with time ($t$) as $\approx t^{-2}$. This leads to an increasing energy deposition in the central region and larger heating of the outer layers.
By about 100 days, the central heating peaks. With time, positrons take over as the main contributors to the energy deposition, and a central `hole' reappears.

 We start with the case of positron transport without $B$. Trapping can be attributed to the opacities only as the positron travels along a ray. 
{ The escape fractions and the normalized total and positron energy redistribution functions are given in Fig. \ref{esca} \& \ref{epositrons}.} 
Up to  about 100 days, positrons continue to be locally confined. At 200 days, the outer layers are 
transparent, diffuse enough that the positrons escape,  lowering the amount high velocity material is heated, but positrons in the inner layers remain locally confined.
At 300 days small features like the dip in Ni at 5,000 km/sec are starting to be washed out, and there is some penetration into the core.  Note that dip in
 $^{56}Ni$ is not physical, but an artifact of spherical DDT. However, it is a good indicator to what degree positron transport has become non-local.
After 500 days, the non-radiating core and the area around it is the only part of the envelope dense enough to have almost complete positron capture. 
At 300 to about 400 days, the  total energy deposition is at a minimum at the center and, subsequently, increases due to the growing mean free path of positrons.

 Magnetic fields determine the path of positrons and thus introduce three more parameters into our discussion: The size of the field, its morphology, and the orientation of the $B$-field relative to an observer.
The main effect of the $B$ field on the escape and energy redistribution can be understood in terms of the Lamour radius $r_L$ compared to the size of the envelope and any sub-structures considered. It increases the 
effective mean free path and influences the direction.  

 For dipole fields, we see that high magnetic fields delay the overall evolution of the energy distribution by about one year. For example, the distribution at 200  
days resembles closely that of the high $B$ field at day 500  (Fig.\ref{epositrons}).

 From the previous sections and within the delayed detonation scenario, we can conclude the following:  
{ Up to about 200 to 300 days, the energy deposition functions do not depend on magnetic fields because $\gamma $-ray 
transport dominate or positrons are deposited locally regardless of the field. Subsequently, positron transport effects may become important, depending on $B$. }

\subsection{NIR Line Profiles for Dipole Fields}

 The goal of this section is to demonstrate the signature of radiation and positron transport  
on late-time spectra. We will discuss in detail the $ 1.644 \mu m$  spectral feature including its evolution with time, 
and its dependence on the orientation of the observer.
 In addition, adjoining  strong  NIR  features are considered to demonstrate the impact of line blending
and how blending may severely limit a kinematic interpretation of profiles. For the spectral band, we have 
calculated a full grid of 30 models for 
$B = 10^{0,3,4,6,9} G$ at  100, 200, 300, 400, 500, 999 days with counters in 5 and 8 directions  
in the Eulerian  $\Theta $ and $\phi$ coordinates. Here, we show only a few of the resulting 1200 
spectra.

 Our reference model, though spherical, requires 
full 3D-transport because of the axial symmetry introduced by the magnetic field.
 Chemical abundances take into account the variability caused by the radioactive decay $^{56}Ni 
\rightarrow ^{56}Co \rightarrow ^{56}Fe$. The excitation balance has been 
calculated using our non-LTE code \citep{h04} as a function of time. For the ionization,  
we found a mean ionization of about 1 to 1.5 electrons per atom, and have assumed a value of 1.5.
We neglect second order effects on the excitation and ionization although 
 the ionization in the central regions will depend  on the heating by positrons
 and at late times, iron group elements may be neutral. As  discussed in the final section, 
this approximation can be tested by observations but can be justified as follows: 
Lower ionization means a slightly larger mean free path, by about 20-30\% (Fig. \ref{pcross}). 
However, a larger mean free path implies more heating
which, in turn, decreases positron transport (see Sect. 2) which, in turn, reduces the feedback effect.
Tests have shown that the total effect on the spectral features remain small.

To discuss line profiles, the continuum fluxes have been subtracted and the maximum fluxes 
normalized. The actual abundances have been used from the model 5p0z22.25.
 For the spectral synthesis, we included the forbidden line profiles of [Fe II] at  
 1.533, 1.560, 1.644, 1.664, 1.677, 1.712, 1.745 $\mu m$ and [CoIII]
of 1.55, 1.74, 1.76 $\mu m$.  
The continuum flux in the $ 1.6-1.7 \mu m$-band are 0.24, 0.15, 0.12,
0.1, 0.06 and 0.02 at 100, 200, 300, 400, 500 and 999 days, respectively.

\begin{figure}
\includegraphics[angle=270,width=0.98\textwidth]{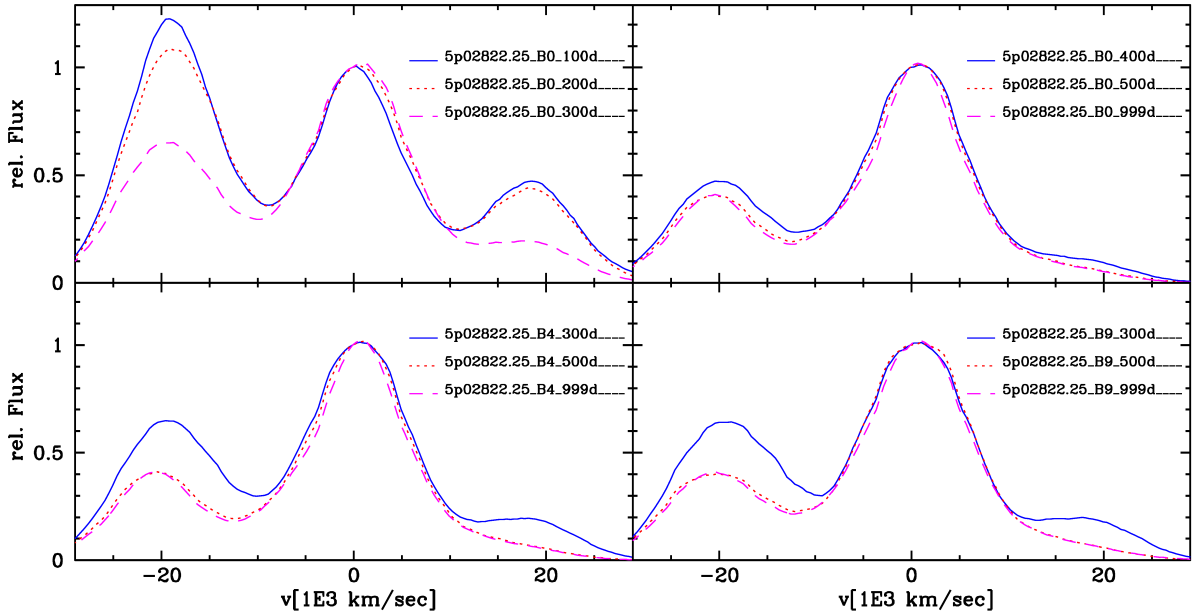}
\caption{  Evolution of the NIR for $B=1 G $, and $10^{4,9} G $ are shown
in the upper and lower panels, respectively. { The continuum flux has been subtracted
and the flux normalized to the peak of the 1.644 $\mu m$ feature. The spectra are labeled
 by $A\_Bb\_Cd\_T\_$ where
$A$ identifies the model , $b=log(B)$, and $C$ is the time after the explosion in days.
$T=1,5$ and $3$ denotes observation from  polar and equatorial direction, respectively,  and $T=\_$ indicates angle averaging.}}
\label{100ave}
\end{figure}

The evolution with time is shown in  Fig. \ref{100ave} in velocity space corresponding to 
wavelengths between 1.5 and 1.8 $\mu m$.
There are three prominent features at about $-20,000 $, $0$ and $+18,000 km/sec$.   

\begin{figure}
\includegraphics[angle=270,width=0.98\textwidth]{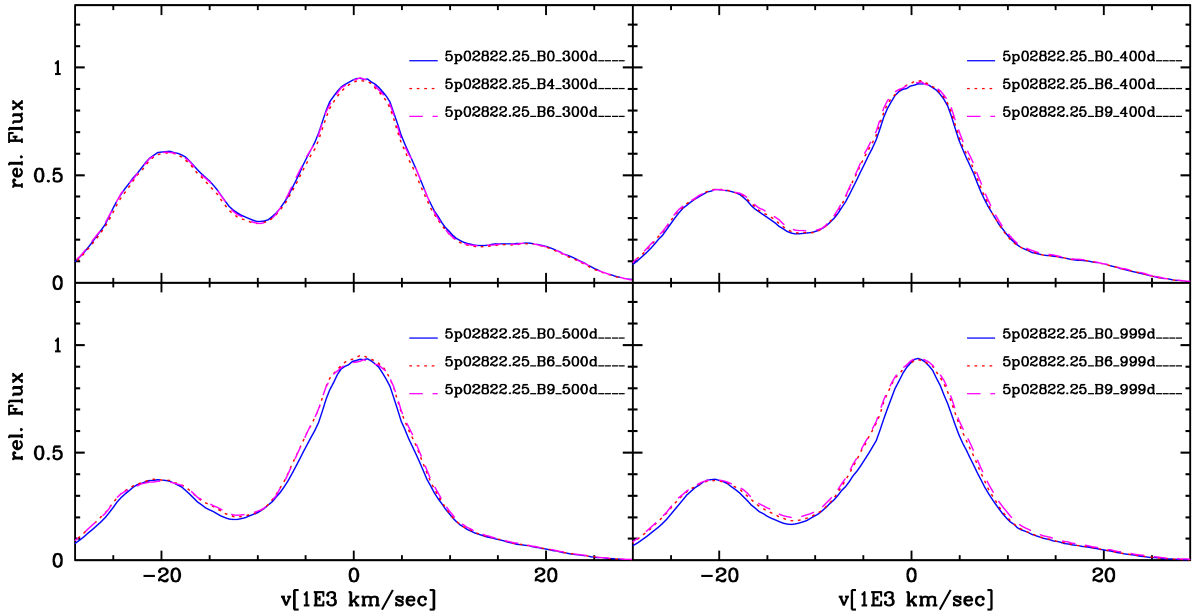}
\caption{Influence of the size of the $B$ field on line profiles
at 300, 400, 500 and 999 days.{ For the labeling scheme, see Fig. \ref{100ave}}.
}
\label{300ave}
\end{figure}

\noindent
{\bf Low Magnetic Fields:}
 We start with a discussion of the angle averaged line profile of [Fe II] at 0 $ km/sec \equiv 1.644 \mu m$
because line-blending effects  for this feature remain small except in the wings \citep{h04}.
By day  100, the profile is peaked because  
gamma-rays are non-local and  mostly absorbed in the central, high density region which has stable $^{54}Fe$.
 The [FeII] feature  is asymmetric and blue-shifted 
 by about 2000 $km/sec$ because the photosphere blocks the redshifted layers of the
envelope.  By day 200, the blue shift is reduced to about $ \approx 800 km/sec$.
 With time, the feature becomes wider and reaches a maximum width at about 300 days.  At around the same time it develops a
`flat-top' or 'stumpy' appearance in the center. { As discussed and shown  in Sect. \ref{DiscussionConclusions},  
a  caustic profile is produced if $^{56}Ni$ is homogeneously mixed  to the center or the explosion of a low-density WD because
the emission would follow the density structure \citep{h04}.
By about day 200, positron transport results in increasingly wider line wings and a small low velocity  
emission component. By day 500, positrons are able to enter and deposit their energy in the core, and the profile 
becomes increasingly peaked.

\noindent
{\bf Role of Magnetic Fields:}  Magnetic fields will hardly change line profiles up to about a year after the explosion.
At about 300 days, the envelope is mostly optically thin in the `quasi-continuum'.
 Observations of the profile of [FeII] at $1.644 \mu m$ at this time provides an excellent tool to measure 
the  distribution of $^{56}Ni$ in the velocity space.
 Earlier, optical depth effects in the continuum and  the contribution of $\gamma $-rays are important contributors  to the
energy input. Later, positrons transport energy into the central regions, and at a rate dependent not 
just on the $^{56}Ni$ distribution and density structure, but also on the magnitude and structure of the $B$-field.
 In practice, early variability in the profiles allows 
us to separate optical depth and geometry effects, to probe the distribution of 
the central non-radioactive Fe and to test for ionization effects. 

 To detect the effects of $B$, we want to focus on the 
late time evolution (Fig. \ref{300ave}). With increasing $B$,
the evolution from 'flat-topped' or 'stubby' to more centrally peaked profiles 
is delayed by about one year for high $B$ fields, as can be expected from the discussion above. Lines are  narrower
by about 20\%  by day 500 compared to models with $B > 10^6 G$. Observations at about 1000 days would allow 
the measurement of  $\approx 10^9 G$.

\noindent
{\bf Perils of Line Blending:} The kinematic interpretation of emission features breaks down if line blending becomes important, 
as  it is the case for features in the optical and  most features in the NIR. 
 The two features neighboring the $1.644 \mu m$ line, at  $-20,000 $ and $+18,000 km/sec$ in 
Fig. \ref{100ave}, are illustrations of the 
perils of apparently clean features. In fact, they are blends.  The former feature is produced by transitions  of [Fe II]  1.533, 1.560 and of  [CoIII] at  1.55 $\mu m$. 
The latter feature is produced by transitions of [FeII]  at  1.712, 1.745 and of [CoIII]  at  
 1.74, 1.76 $\mu m$.   

The evolution of both these `satellite' features is dominated by the decay of radioactive
$^{56}Co$. Up to about day 200, the forbidden $^{56}Co$ lines are responsible for
most of the emission. After about day 300, only $^{56}Fe$ is left, as evidenced also
by the  slowly varying line ratio between the 1.644 $\mu m$ feature and the 'satellites'.

{Early on, the red feature at $+18,000 km/sec$  shows a blueshift similar to [Fe II] at 1.644 $\mu m$ but it 
is broader because with several line transitions contribute including [Fe II] at 1.712,  1.745 $\mu m$, and  [CoIII] $ 1.74, 1.76 \mu m$. 
 One may expect that, with time, the feature would be shifted towards the blue because of the longer wavelength of the [Co III] transitions.
However, we see a red-shift because the feature formed on the wing of the 1.644 $ \mu m$ which produces a 'sloped' quasi-continuum.} 
  
 {  The  $-20,000 km/sec$ feature at $\approx 1.55 \mu m$ dominates the NIR spectra early on. However,  due to its nature, i simple interpretation may be
misleading.   With time, this feature undergoes a blue shift opposite to
the behavior of the [Fe II] at 1.644 $\mu m$. Taking at face value and in a kinematic 
interpretation, this would mean that we see an off-set in $^{56} Ni $ whereas this shift, at least 
in the models, is  due to the changing importance of [Co III] vs. [FeII]. 

In summary, time evolution allows us to detect and, potentially, to separate line blending,
geometry and $B$ field effects. The [Fe II] feature at $1.644 \mu$ is rather unique in the NIR. }

\noindent
{\bf Directional Dependence of NIR Features:}
Large-scale magnetic dipole fields introduce  a directional dependence of the observed  
line profile. The variations in the profiles can be understood in terms of the Lamour radius
given in Tab. \ref{table1}.
 We consider only times later than 300 days when positron  transport effects may be  significant.
 For low $B$ fields, the Lamour radius is much larger than the envelope and positrons
travel on rays which results in isotropic profiles.

\begin{figure}
\includegraphics[angle=270,width=0.98\textwidth]{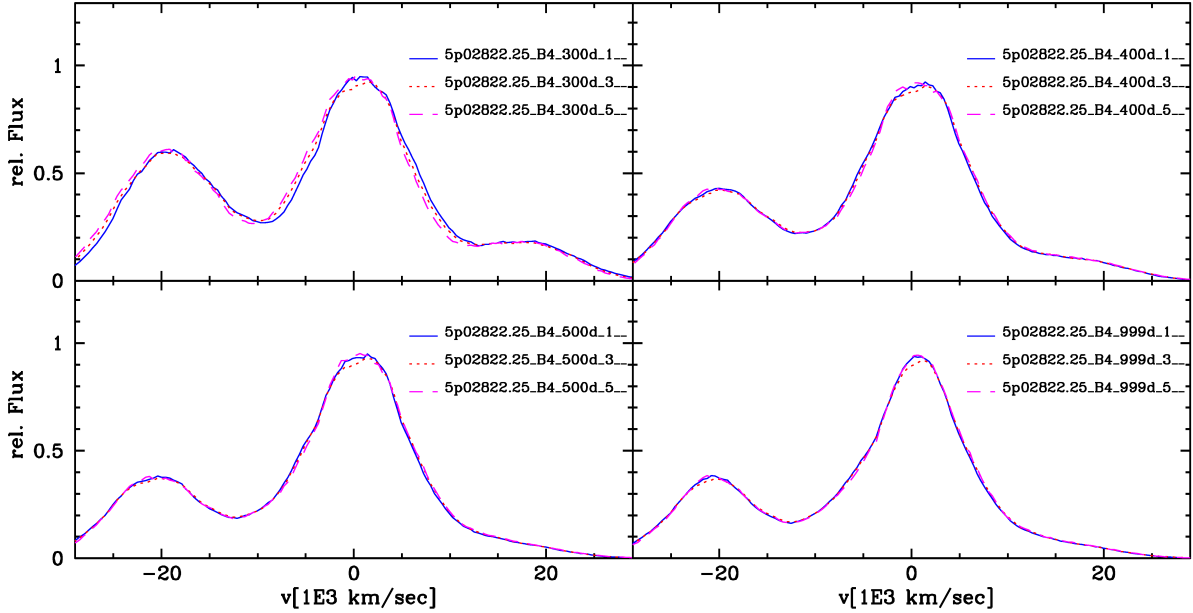}
\caption{Directional dependence of the spectrum of the
reference model as a function of the Doppler shift
relative to the forbidden [FeII] at 1.644 $\mu m$.
assuming a dipole field $10^4 G$. The solid, dotted and
slashed lines correspond to $\Theta $ of 0, 90 and 180
degrees. The panels show the spectra at 300, 400, 500
and 999 days after the explosion.
}
\label{proB4}
\end{figure}

\begin{figure}
\includegraphics[angle=270,width=0.98\textwidth]{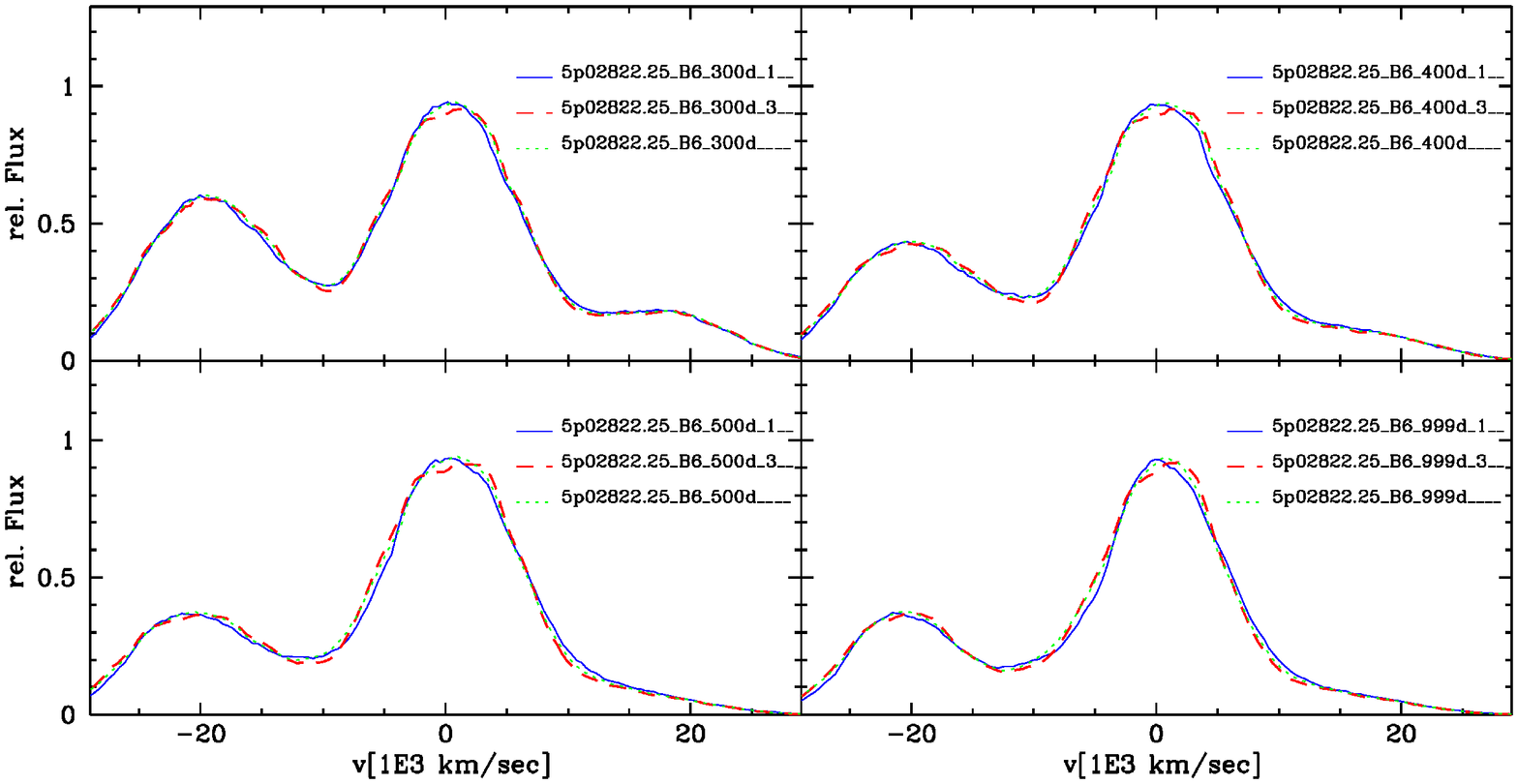}
\caption{Same as fig. \ref{proB4} but $B=10^6 G$. 
}
\label{proB6}
\end{figure}  

\begin{figure}
\includegraphics[angle=270,width=0.98\textwidth]{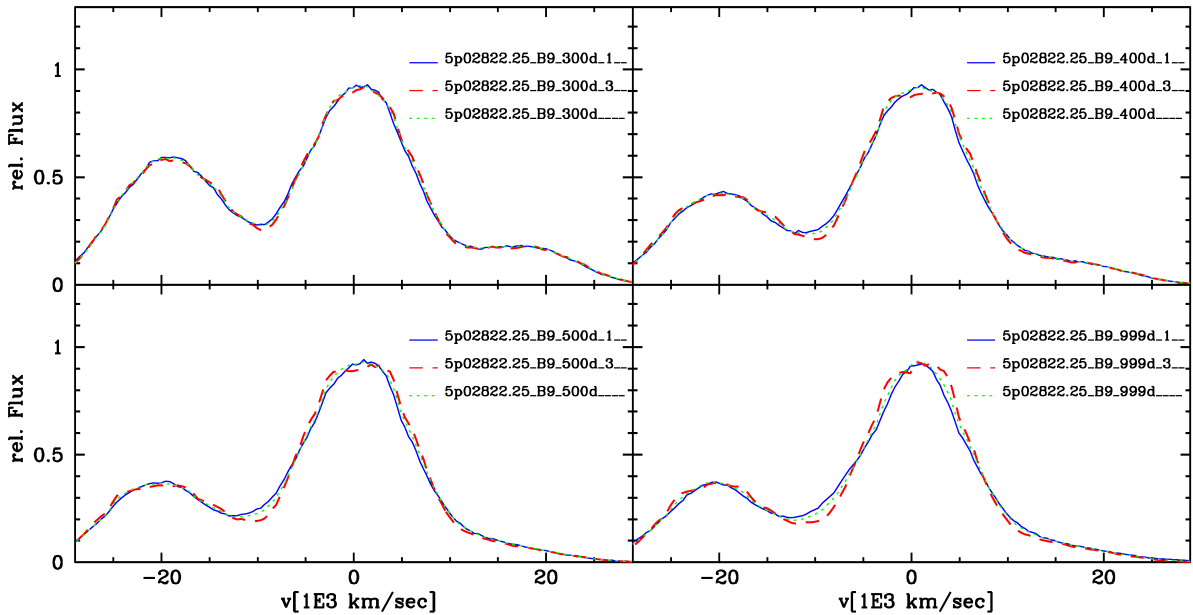}
\caption{Same as fig. \ref{proB4} but $B=10^9 G$. 
}
\label{proB9}
\end{figure}

 In Figs. \ref{proB6} \& \ref{proB9}, the directional dependence of profiles are shown for high magnetic fields.
For $B=10^{9}G$, the positrons follow the field lines with an increase in the 'effective' cross section. Variations 
in the  energy deposition and, thus, the energy deposition depends mainly on the mean free path.
 By day 1000, positrons created in the polar regions can travel to the center regardless of the $B$ field.
More generally, positrons created at more equatorial latitudes are funneled towards the pole.
This geometry effect causes maximum directional dependence. From the pole, we see flat-topped profiles with 
an increased line width. Seen from the equator, late time profiles are centrally peaked.  

 Lowering the $B$ field increases the mean free path, and the path of the positron is bent but 
does not exactly follow the field lines. For $B=10^6G$, the directional dependence shows up earlier.
 High precision observations of the time dependence of profiles would allow us to learn about the orientation
of the field relative to the observer.

\subsection{Influence of the Morphology of the Magnetic Field}
\label{morphoc}

This leads us to the effects of the morphology of the $B$ fields. We may expect a dipole field in a WD
prior to the runaway but it may be distorted by the turbulent motion during the thermonuclear runaway, or by RT-instabilities during the explosion of the WD. 

 As an example and limiting case we have calculated the transport for a turbulent field.
In calculations of the thermonuclear runaway, the largest turbulent eddies are
$\approx 5\%$ of the WD \citep{hs02}. We note that the energy deposition is close to isotropic. 

The results are dominated by the change of the gyro-radius relative to the size of the structure of the $B$ field (Tab. \ref{table1}). $B$ in excess of $10^6 G$ results
in local trapping of positrons with some transport effects starting at about 500 to 1000 
days. For fields less than $\approx 10^4 G$, the gyro-radius is much larger than the
size of the envelope and, consequently, the results are identical to the angle-averaged 
profiles in Figs. \ref{100ave} \& \ref{300ave}. Probing the magnetic fields require time series of
profiles after about 2 to 3 years, or magnetic fields between $10^{4...6}G$ observed 
beyond 400 days. { As discussed above for very high magnetic fields, LCs come close to the case of full trapping}.

\section{Discussion and Conclusions}\label{DiscussionConclusions}

We have presented a new implementation of positron transport using a Monte Carlo method. This allows us to the study of 
transport effects in full 3D for arbitrary magnetic fields without the need for either the low or large field approximations.
                                                                                                                                               
 In SNe~Ia, light curves and spectra are powered by the radioactive decay from $^{56}Ni \rightarrow ^{56}Co  \rightarrow ^{56}Fe$. 
The evolution of the energy deposition is dominated by a transition from the $\gamma $-ray to the positron dominated regime between 
200 and 300 days, and the changing optical depth due to the expansion.
 Because positron transport depends on $B$, late time observations provide a tool to probe the fields.

 Our study is based on the explosion of a $M_{Ch}$ mass WD, specifically a delayed detonation model for a Branch-normal SN~Ia, with
$B$ fields imprinted at the time of the explosion.  Up to about 200 days after the explosion, positron transport can be neglected.
 By 300 days, positron transport becomes important and changes both the total and redistribution of energy from radioactive decay. 
 
{ 
  The effect of $B$ on the late time light curve grows from $\approx 0.1 ^m$ at day 300 to $ \approx 1^m$ 
some three years after the explosion.  We found LC variations due to $B$ comparable to previous studies within 
$\approx 20 \%$ \citep{milne99}. Differences can be attributed to the magnetic field structure assumed. 
Milne and previous authors considered radial magnetic fields whereas we assumed that the $B$-field is frozen into the 
expanding ejecta. Our assumption is justified because the expansion is homologous. The additional component in $B$ results in a larger degree 
of trapping. In practice, however, the use of LCs may be limited. It relies on monochromatic  V and UBVRI LCs
to reconstruct the bolometric LC (Sect. \ref{Results}). 
 Photon redistribution from the visual to the mid- and far-IR becomes increasingly important starting several months after the explosion  and, after a few years,
may result in an 'IR-catastrophe'.  An additional problem may be related to the fact that LCs measure the total energy input.
 Additional energy contributors may become important, such as interaction with the circumstellar medium. Moreover, the diversity in SNe~Ia
may further limit the use of differential analysis based on a single quantity.

 The use of late-time line profiles of iron group elements overcomes some of these problems. 
The formation of these lines depends on the population of the upper atomic level and, thus, profiles are hardly effected  by 
the overall photon redistribution or e.g. ongoing interactions which deposit energy in the outer  rather than the $^{56}Ni$ layers.   
 Line profiles are a measure of the redistribution function of the input energy in velocity space rather than the escape fraction. 

  We have demonstrated that magnetic fields can be probed by the line shape and its evolution.  
 Line blending due to Doppler shift is a major problem in SNe~Ia but we have identified as a suitable spectral feature the forbidden [Fe II] line at $1.644 \mu m$.  
  This feature shows only minor blending and thus allows a kinematic 
 interpretation of the profile if optical depth effects are taken into account. Other strong, individual NIR 
 (and optical) features are of limited use  because they originate from strong blends, mostly of [Fe II] and [CoIII] lines.

 We find that large scale, dipole fields have the biggest effect on the line profile.
Up to about 200-250 days after the explosion, positrons are mostly locally trapped and, consequently,
the profile depends on the chemical and density structure of the explosion models. The change with 
time of the feature can be understood by the change from the regime of energy input by $\gamma $-rays to
that of positrons. Line profiles depend on the geometry and distribution of $^{56}Ni$ and the density structure of 
the explosion model \citep{h04,motohara06,hoeflich06,maeda11}. At early times, profiles are peaked because of 
excitation due to non-local $\gamma $-rays which, with time, become increasingly 'flat-topped' until non-local 
positrons can excite the low velocity iron-group elements.}

 To produce 'flat-topped' profiles at day 300-400 as have been observed in SN2003du and SN2004hv, 
we need magnetic fields of $10^6 G$ or larger. For large-scale dipole fields, the supernovae must be seen from equatorial 
directions. Seen from an angle close to the pole, the profile is broad ('stumpy'), but with a 'round-top' because positrons 
can travel towards the central region parallel to the magnetic field. 

To separate optical depth effects of the envelope, ionization and orientation effects, time 
series of the NIR are crucial. Spectra are needed with a resolution of, at least, 300 to 500 km/sec  
between 300 to 400 days, and a signal to noise ratio of 10 to 20 \% ,  or a signal to noise of 20 to 
30 \% after 500 days.

{ 
We discussed the influence of the morphology of the B-field. 
For magnetic fields of turbulent scales, the profiles will be non-directionally dependent, and  somewhat smaller 
$B$ fields are sufficient.  To separate various morphologies of $B$, the typical mean free path of positrons must be larger than the 
characteristic structures, which would require spectra beyond day 400. These effects are currently under investigation.
 
  Our simulations suggest the need of $B$ in excess of $ 10^{6} G$ to explain
persistent 'flat-topped' or 'stumpy' profiles as observed for SN2003du and SN2003hv  after about a year.
 As discussed below and in \citet{h13}, fields of this size will affect the hydrodynamics of the 
smoldering phase of the WD just prior to the runaway and, possibly, deflagration burning. 
This is the first strong, observational evidence that magneto-hydrodynamical effects should be taken into account 
during the runaway and deflagration phase of SNe~Ia.
 We also note that explosion models with central densities less than $10^9 g/cm^3$ will give $^{56}Ni$ up to the 
center and produce narrow overall profiles consistent with the 'caustic' density profile of SNe~Ia (see Fig.\ref{chem}) 
as discussed in \citet{h04,stritzinger14}. Thus, SN2003du and SN2003hv are likely to originate from $M_{Ch}$ mass explosions
with very limited amount of mixing of $^{56}Ni $ during the explosion.}

{ However, SNe~Ia are a diverse group of objects as discussed in the introduction. Positron
transport effects depend on the free mean path, thus, the density and $B$.  The time of 
importance will depend on the overall $^{56}Ni $ distribution. In Fig. \ref{proce}, we show the evolution of our
reference model but with central $^{56}Ni$.  The late time profiles are 'peaked' rather than 
'flat-topped' or 'stubby' unlike our reference model. 
 Because of the higher densities, positron transport effects remain small and evolution starts at about 3 years
rather than 1 to 1.5 years.  The [Fe II] feature shows little evolution between 
200 days and about 3 years after the explosion. This kind of profiles may be produced either by 
lower central densities of the WD and/or mixing during the explosion.
 The distribution of $^{56}Ni$ is important and line profiles will allow us to decipher both $B$ fields and
the structure.

\begin{figure}
\includegraphics[angle=360,width=0.98\textwidth]{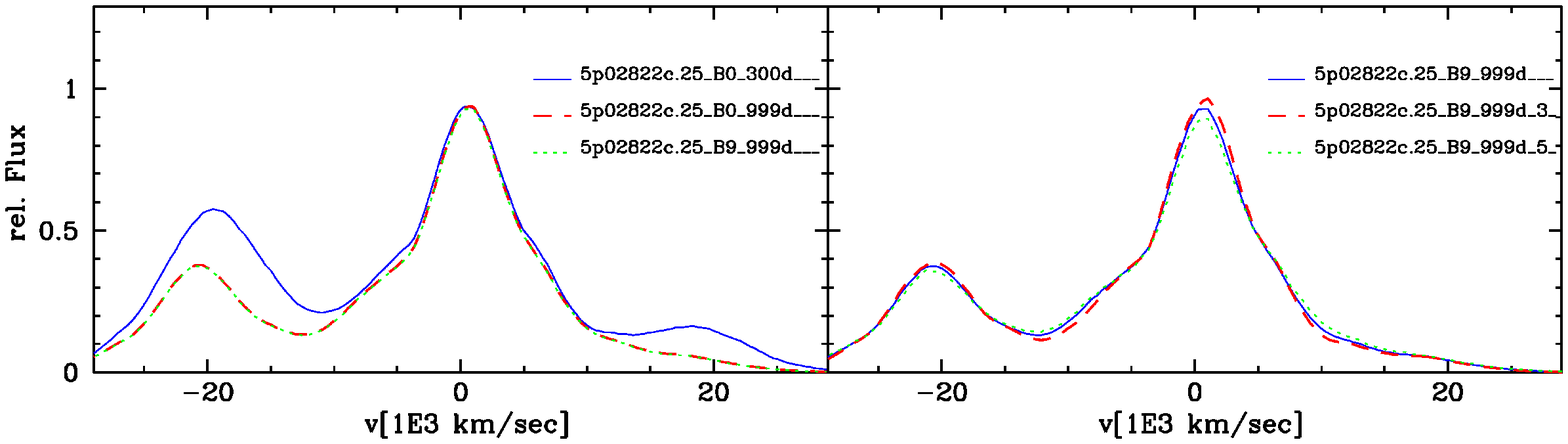}
\caption{{ NIR profile of our reference model but assuming homogeneous
mixing of $^{56}Ni$ down to the center. The average profile of the 1.644  $\mu m$ feature shows
little evolution between 300 and 999 days other than due to the
blends in the line wings (left). The profile is peaked rather than 'flat topped' or 
'stumpy'. On the right, the average profile is compared to the polar and equatorial 
profiles for $10^{9}G$.  By day 999 and high magnetic and dipole fields, the profiles 
have directional dependence.}  
}
\label{proce}
\end{figure}

 In this context, SN~2014J in M82 \citet{Fossey14} may shed new light on SNe~Ia.
 It is the first SNe~Ia for which $\gamma $ rays have been detected. These are consistent
with a $^{56}Ni$ production of $0.6 M_\odot$ \citep{Isern14,Churazov14}. 
 Recently, \citep{diehl14} analysed the early $^{56}Ni$ lines at 158 and 812 MeV
in SN~2014J. The authors report $^{56}Ni$ lines at 158 and 812 MeV 
 with a line width of $\approx 5000 km/sec$ and which are not shifted in rest-frame of the supernovae. The line fluxes 
 require about $0.06 M_\odot$ of $^{56}Ni$. 
 \citep{diehl14} suggest a configuration in which helium is accreted in a belt
around the progenitor seen 'pole-on'. During the explosion, parts of the He may be burned to $^{56}Ni$. 
If confirmed, this detection clearly demonstrates the diversity in $^{56}Ni$ distributions.
 In framework of this study, we may expect a narrow component on top of broader un-blended features 
 in late-time NIR and MIR spectra. Because of the lower densities, we expect positron transport 
effects starting earlier for the high velocity component at about 100 to 200 days depending on $B$. 
For SN2014J, corresponding observations during the next few years are scheduled which will allow a S/N at the level required. 
Note that an $0.1^m$ effect on the bolometric is unlikely to be detected by light curves.}

Finally, we have to emphasize the limitations of this study: 
we neglected the feedback of positron transport on the ionization structure. Namely, we may have 
central neutral iron group elements \citet{liu97,jeffery98,liu98} which can produce flat-topped profiles 
as ionization effect. However, the location of  ionization fronts will be time-dependent and hardly produce
persistent characteristics.
 
 Our study suggests high initial  magnetic fields and $M_{Ch}$ WD for several SNe~Ia. However, we lack 
time-series needed for more detailed analyses. Currently,  a time series of SN2005df is being analyzed 
as part of the PhD thesis of Tiara Diamond.  Obviously, we  need more and time sequences in the NIR and this kind
of observations will become more common in the near future.
{ We will observe the nearby SN~2014J over the next 
few years to obtain S/N of a few percent, and corresponding are scheduled at Gemini. JWST, GMT, ELT and
WFIRST will allow us to obtain NIR spectra on a similar or better S/N level for SNe~Ia up to about 
Coma distance.}

 For turbulent magnetic fields, the quantitative results will depend on the turbulence spectrum present in the 
 WD or created during the deflagration phase of burning. Currently, we study the influence of more  realistic 
 turbulent fields, and their influence on line  profiles.
Another question is related to the origin of high $B$ fields.
WDs are observed with $B$-fields up to several times $10^7$~G, but the
majority have no measurable B field \citep{liebert03,schmidt03,Silvestri07,tout08}.
 Large $B$-fields may be produced either during the accretion phase, right before the thermonuclear 
 runaway, or the deflagration phase. 
In a related project, we are currently studying properties of magnetic fields in the thermonuclear
runaway: the growth of the field by dynamo action; impact of magnetic fields on
velocity statistics; the suppression of the Rayleigh-Taylor instability by magnetic tension; 
and the alteration of flame speeds by anisotropic conduction. 
 In three dimensional studies, there is a tendency for magnetic fields to
increase the rate of growth in modes parallel to the field, due to the
suppression of secondary instabilities, and a decrease in growth in modes
transverse to the magnetic field \citep{Stone07,Stone07b, Ghezzi04} which, eventually, may 
lead to an answer why we do not see RT instabilities to the extent predicted. 

{  Above, we have discussed the limitations of light curve with respect to $B$. A combination of 
line profiles and LCs may help to overcome some of the problems with using LCs, in particular, if combined with
observations of SNe~Ia in the mid-IR to calibrate the photon redistribution effects. We are currently extending our LC studies along this path. }

\acknowledgments
We would like to thank many colleges and collaborators for their support, in particular, E. Baron, D. Collins, T. Diamond, R. Fesen, A. Hamilton, E. Hsiao, A. Khokhlov, J. Maund, M. Phillips, D. Rubin, M. Stritzinger, N. Suntzeff, C. Telesco, S. Valenti, L. Wang, J.C. Wheeler and many observers to provide insights and data. The work presented in this paper has been supported by
the NSF projects AST-0708855, ``Three-Dimensional Simulations of Type Ia Supernovae: Constraining
Models with Observations'' and AST-1008962, ``Interaction of Type Ia Supernovae with their Environment".
In parts, the results presented have been obtained in course of a PhD thesis by
 R. Penney at Florida State University.

\bibliography{article}

\end{document}